\documentclass[aps,pre,twocolumn,superscriptaddress,10pt,showpacs]{revtex4}
\usepackage[dvips]{graphics}
\usepackage{graphicx}
\usepackage{amsfonts}
\usepackage{amssymb}
\usepackage{amsmath}
\usepackage{subfigure}

\begin{document}

\title{Discrete surface solitons in two dimensions}
\author{H.\ Susanto}
\affiliation{Department of Mathematics and Statistics, University of Massachusetts,
Amherst MA 01003-4515, USA}
\author{P. G.\ Kevrekidis}
\affiliation{Department of Mathematics and Statistics, University of Massachusetts,
Amherst MA 01003-4515, USA}
\author{B. A.\ Malomed}
\affiliation{Department of Interdisciplinary Studies, School of Electrical Engineering,
Faculty of Engineering, Tel Aviv University, Tel Aviv 69978, Israel}
\author{R.\ Carretero-Gonz\'alez}
\affiliation{Nonlinear Dynamical Systems Group, Department of Mathematics and Statistics,
and Computational Science Research Center, San Diego State University, San
Diego CA, 92182-7720, USA}
\author{D. J.\ Frantzeskakis}
\affiliation{Department of Physics, University of Athens, Panepistimiopolis, Zografos,
Athens 15784, Greece }

\pacs{05.45.Yv, 03.75.-b, 42.65.Tg}

\date{Submitted to {\em Phys.~Rev.~E}, July 2006}

\begin{abstract}
We
investigate fundamental localized modes in 2D lattices with an edge (surface).
Interaction with the edge expands the stability area for
ordinary solitons, and induces a difference between perpendicular and
parallel dipoles; on the contrary, lattice vortices cannot
exist too close to the border.
Furthermore, we show analytically and numerically
that the edge stabilizes a novel wave species,
which is entirely unstable in the uniform lattice,
namely, a ``horseshoe'' soliton, consisting of 3 sites.
Unstable horseshoes transform themselves into a pair
of ordinary solitons.
\end{abstract}

\maketitle





\section{Introduction and the model}
Solitons on surfaces of fluids \cite%
{book}, solids \cite{Gerard}, and plasmas \cite{StenfloPlasma}
are classical objects of experimental and theoretical studies
of nonlinear science. Recently, a new
implementation of surface solitary waves
was proposed \cite{DemetriPrediction} and experimentally created \cite%
{Demetri} in nonlinear optics, in the form of \textit{discrete} localized
pulses supported at the edge of a semi-infinite array of nonlinear
waveguides. Two-component surface solitons were analyzed too \cite{vectorial}%
, and it was predicted that solitons may be supported at an edge of a
discrete chain by a nonlinear impurity \cite{Molina}. Parallel to that,
surface gap solitons were predicted \cite{BarcelonaGS} and created in an
experiment \cite{experiment} at an edge of a waveguiding array built into in
a self-defocusing continuous medium. Very recently, the
experimental creation of
discrete surface solitons supported by the quadratic nonlinearity was
reported as well \cite{chi2}. In these cases, the solitons are one-dimensional
(1D). In Ref.~\cite{BarcelonaCrossVortexSaturable}, a 2D medium with
saturable nonlinearity was considered, with an embedded square lattice, that
has a jump at an internal interface;
there, stable asymmetric vortex solitons
crossing the interface were predicted,
as a generalization of discrete vortices on 2D lattices \cite{we} and vortex solitons supported
by optically induced lattices in
photorefractive media \cite{photorefr}.
Solitons supported by a nonlinear defect at the edge of a 2D lattice were also
considered
\cite{Molina2D}.

The search for surface solitons in lattice settings is
a
natural problem, as, in any experimental setup, the lattice inevitably has an edge.
In this paper, we report new
results for surface solitons in semi-infinite 2D lattices. We will
first consider straightforward generalizations of localized modes
studied previously on uniform lattices, viz., fundamental solitons
and two types of dipoles, oriented perpendicular or parallel to
the surface. Then, we will introduce a novel species, the so-called
\textit{horseshoe soliton}, in the
form of an arc abutting upon the lattice's edge. The existence,
and especially the stability,
of such a localized mode is a nontrivial issue, as attempts to find
a ``horseshoe'' in continuum media with imprinted lattices and an
internal interface
(similar to the medium
in Ref.~\cite{BarcelonaCrossVortexSaturable})
have produced negative results \cite{Yaroslav}.
We find that, in
the semi-infinite discrete medium, a horseshoe
not only exists near the lattice edge, but perhaps more importantly
has a stability region. For comparison, we also construct a family of
the same wave pattern in the uniform lattice [which is a form of a
stationary localized solution of the 2D discrete nonlinear Schr\"{o}dinger
(DNLS) equation that has not been discussed previously and we
illustrate how/why it is of interest in its own right]. In particular,
we find that this family of solutions is \emph{completely unstable}
in the bulk of the uniform lattice,
which stresses the nontrivial character of the surface-trapped horseshoes,
in that they may be stabilized (for an appropriate parameter range)
by the lattice edge.

The model of a semi-infinite 2D array of waveguides with a horizontal edge,
that we consider below, is based on the DNLS equation for wave amplitudes $%
u_{m,n}(z)$ in the guiding cores, $z$ being the propagation distance:
\begin{eqnarray}
iu_{m,n}^{\prime }&+&C
(u_{m+1,n}+u_{m-1,n}+u_{m,n+1} \nonumber \\[1.0ex]
&+&u_{m,n-1}-4u_{m,n})
+|u_{m,n}|^{2}u_{m,n}=0,
\label{bulk}
\end{eqnarray}
for $n\geq 2$ and all $m$, where the prime stands for $d/dz$, and $C$ is the
coupling constant. At the surface row, $n=1$, Eq.~(\ref{bulk}) is modified
by dropping the fourth term in the above parenthesis of Eq. (\ref{bulk})
(cf.~the 1D model in
Refs.~\cite{Demetri}),
namely, $u_{m,0}=0$ as there are no waveguides at $n\leq 0$.
Note that, despite the presence of the edge, Eq.~(\ref%
{bulk}) admits the usual Hamiltonian representation, and conserves the total
power (norm)\thinspace, $P=\sum_{m=-\infty }^{+\infty }\sum_{n=1}^{+\infty }\left\vert u_{m,n}\right\vert ^{2}$.
%
%

Stationary solutions to Eq.~(\ref{bulk}) will be looked for as
$u_{m,n}=e^{ikz}v_{m,n}$, where the wavenumber $k$ may be scaled to $1$, once $%
C$ is an arbitrary parameter, and the stationary solution obeys the
equation
\begin{eqnarray}
(1&-&|v_{m,n}|^{2})v_{m,n}-C(v_{m,n+1}+v_{m,n-1} \nonumber \\[1.0ex]
&+&v_{m+1,n}+v_{m-1,n}-4v_{m,n})=0,
\label{f}
\end{eqnarray}
with the same modification as above at $n=1$.

We will first report results of an analytical approximation for
the shape and stability of dipoles and ``horseshoes", 
valid for a weakly coupled lattice ($C\rightarrow 0$). This will
be followed by presentation of corresponding numerical results.
Finally, we
will briefly discuss the interaction of
vortices with the lattice's edge.

\section{Perturbation analysis}
Analytical results can be obtained for small
$C$, starting from the anti-continuum (AC)\ limit, $C=0$ (see Ref.~\cite%
{peli05} and references therein). In this case, solutions to Eq.~(\ref{f})
may be constructed as a perturbative expansion
$$
v_{m,n}=\sum_{k=0}^{\infty }C^{k}v_{m,n}^{(k)}.
$$
In the AC limit proper, the \textit{seed solution}, $%
v_{m,n}^{(0)}$, is
zero except at a few \textit{excited sites},
which determine the configuration.

By means of the analytical method, we will consider the following
configurations: (a) a fundamental surface soliton, seeded by a single
excited site, $v_{1,1}^{(0)}=1$
(the first subscript $1$ denotes the soliton's location in the horizontal direction),
(b) surface dipoles, oriented perpendicular (b1) or
parallel (b2) to the edge, each seeded at two sites,
\begin{equation}
\left\{ v_{0,1}^{(0)},v_{0,2}^{(0)}\right\} =\left\{ -1,1\right\} ,~\mathrm{%
or~}\left\{ v_{0,1}^{(0)},v_{1,1}^{(0)}\right\} =\left\{-1,1\right\} ,
\label{dipole}
\end{equation}
and (c) the ``horseshoe'' 3-site structure
\begin{equation}
\left\{ v_{1,1}^{(0)},v_{0,2}^{(0)},v_{-1,1}^{(0)}\right\} =\left\{
e^{i\theta _{1,1}},e^{i\theta _{0,2}},e^{i\theta _{-1,1}}\right\} ,
\label{stemcell2}
\end{equation}
with $\theta _{1,1}=0,\,\theta _{0,2}=\pi ,\,\theta _{-1,0}=2\pi $.
We note in passing that
stable dipole states on the infinite lattice were predicted in Ref.~\cite%
{Bishop}, and later observed experimentally
in a photorefractive crystal
\cite{dipole}. All the above seed
configurations are real, and the horseshoe may, in principle,
also be regarded as a truncated quadrupole, which is a real solution as well
\cite{we2}.

At small $C>0$, it is straightforward to calculate corrections to the
stationary states at the zeroth and first order in $C$.
Then, the stability of each state is determined by a set of eigenvalues, $\lambda$,
which are expressed in terms of eigenvalues $\mu $ of the \textit{%
Jacobian matrix}, to be derived in a perturbative form, $\mathcal{M}%
=\sum_{k=0}^{\infty }C^{k}\mathcal{M}_{k}$, from the linearized equations for
small perturbations around a given stationary state \cite{peli05}. The
stability condition is $\Re(\lambda ) = 0$ for all $\lambda$,
for the Hamiltonian system
of interest herein (since if $\lambda$ is an eigenvalue, so are
$-\lambda$, $\lambda^{\star}$ and $-\lambda^{\star}$, where the
asterisk denotes complex conjugation).

For the dipole and horseshoe configurations, (b1,b2) and (c), the
calculations result in
\begin{eqnarray}
\mathcal{M}^{(b)}&=&C\left(
\begin{array}{rr}
-1 & 1 \\
1 & -1%
\end{array}
\right) +\mathcal{O} \left(C^{2}\right), \nonumber \\[3.0ex]
\mathcal{M}^{(c)}&=&C^{2}\left(
\begin{array}{rrr}
-4 & 2 & 2 \\
2 & -1 & -1 \\
2 & -1 & -1%
\end{array}%
\right) +\mathcal{O} \left(C^{3}\right) ~
\end{eqnarray}
(the matrices for (b1) and (b2) coincide, at this order). From here, we obtain
\emph{stable} eigenvalues, $\lambda _{1}^{(b)}=0,\lambda _{2}^{(b)}=\pm 2%
\sqrt{C}i+\mathcal{O}\left( C\right) $, and $\lambda _{1}^{(c)}=0,\lambda
_{2}^{(c)}=\mathcal{O}\left( C^{2}\right) ,\lambda _{3}^{(c)}=\pm 2\sqrt{3}%
Ci+\mathcal{O}\left( C^{2}\right) $. Both for the dipoles and half-vortex,
one eigenvalue is exactly zero, as this corresponds to the Goldstone mode
generated by the phase invariance of the underlying DNLS\ equation. As for
eigenvalue $\lambda _{2}^{(c)}$, it becomes different from zero at order $%
\mathcal{O}\left( C^{2}\right) $, and, as shown below, it plays a critical
role in determining the stability of the horseshoe structure.

We do not consider here the fundamental soliton, (a),
as its destabilization mechanism is different
(and requires a different analysis)
from that of the dipoles and horseshoes;
in particular, the critical eigenvalues bifurcate not from zero, but
from the edge of continuous spectrum
(see below).

\begin{figure}[tb]
\centerline{
\includegraphics[width=4.4cm,angle=0,clip]{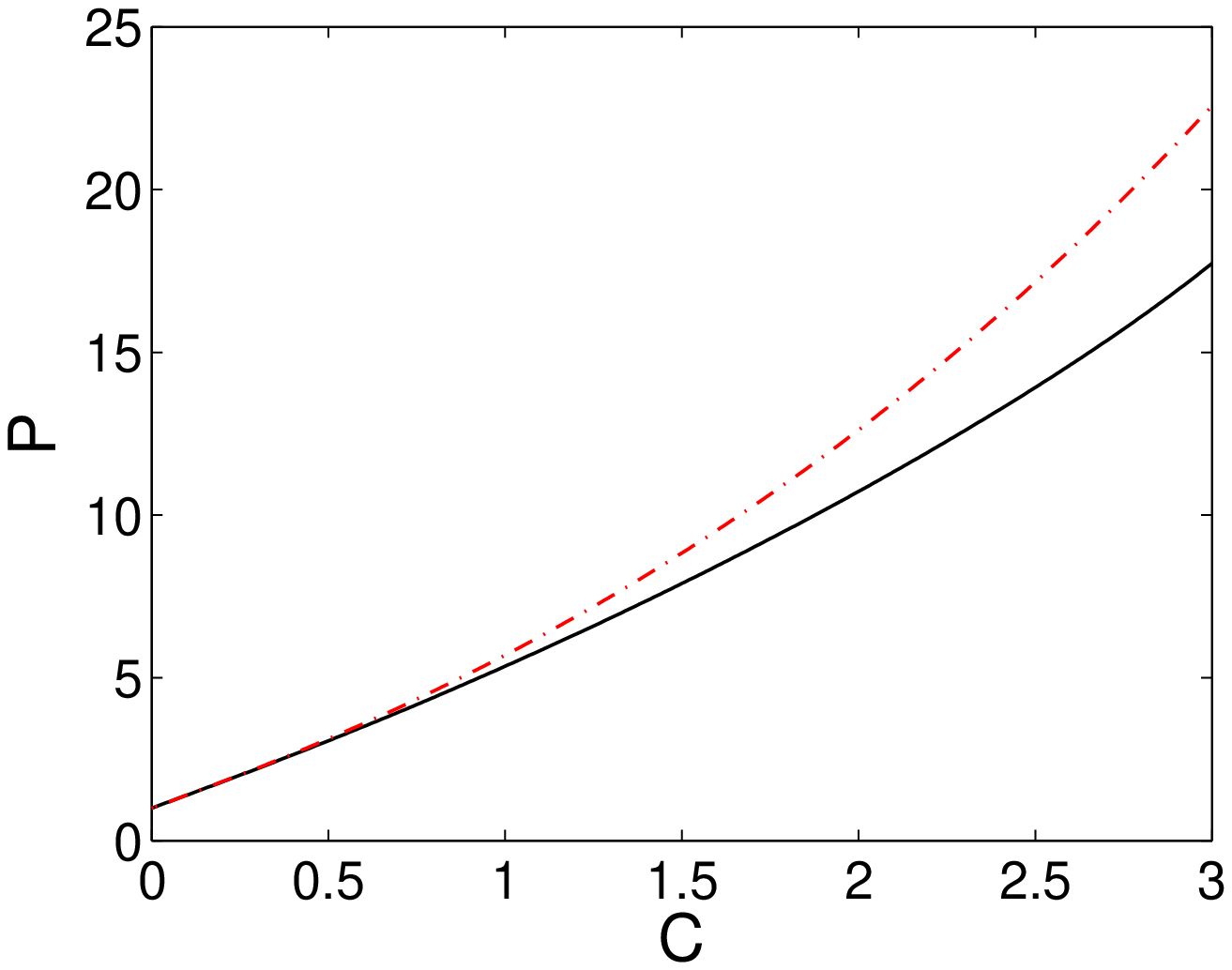}
\hskip-0.2cm
\includegraphics[width=4.4cm,angle=0,clip]{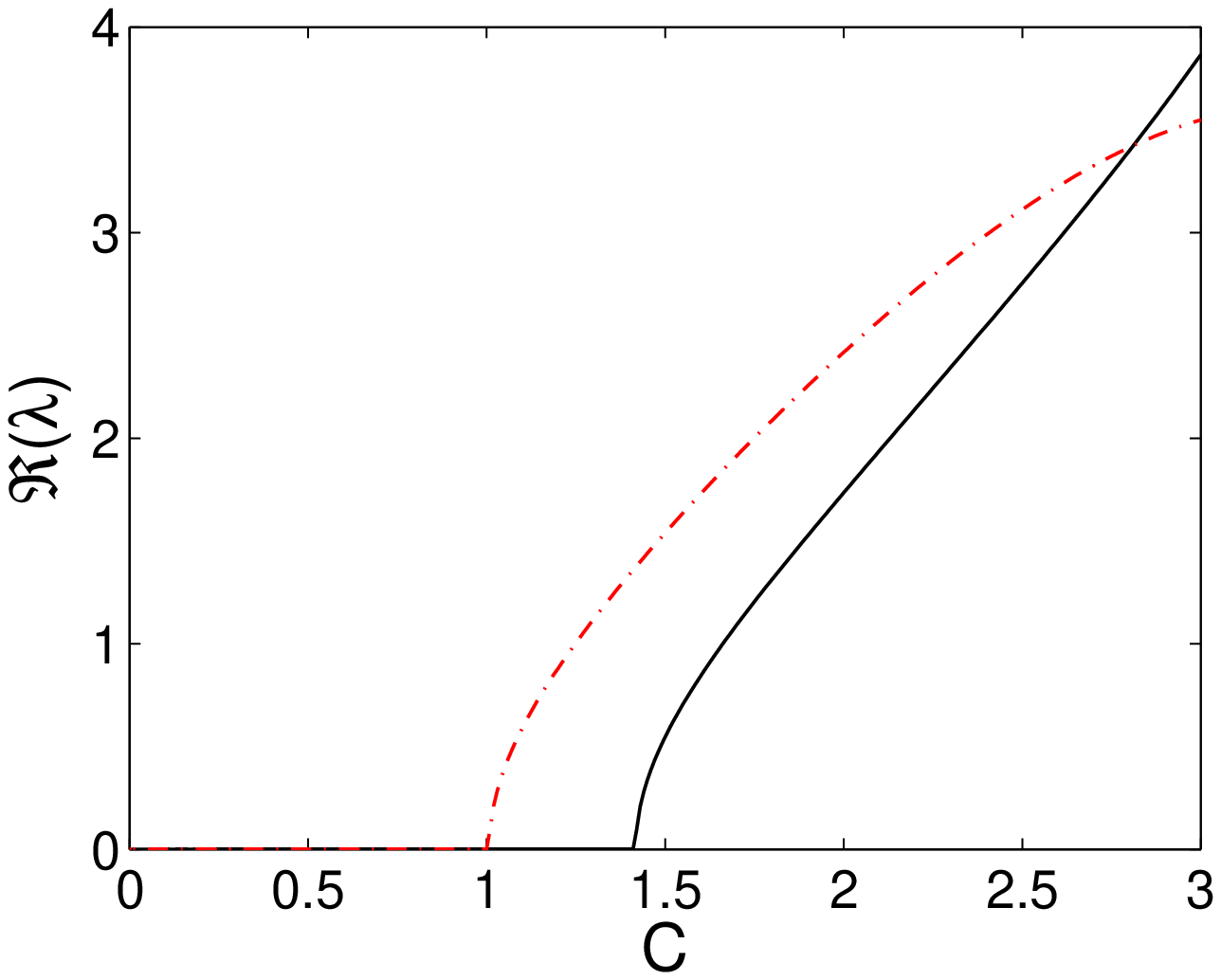}
}
\centerline{
\includegraphics[width=4.4cm,angle=0,clip]{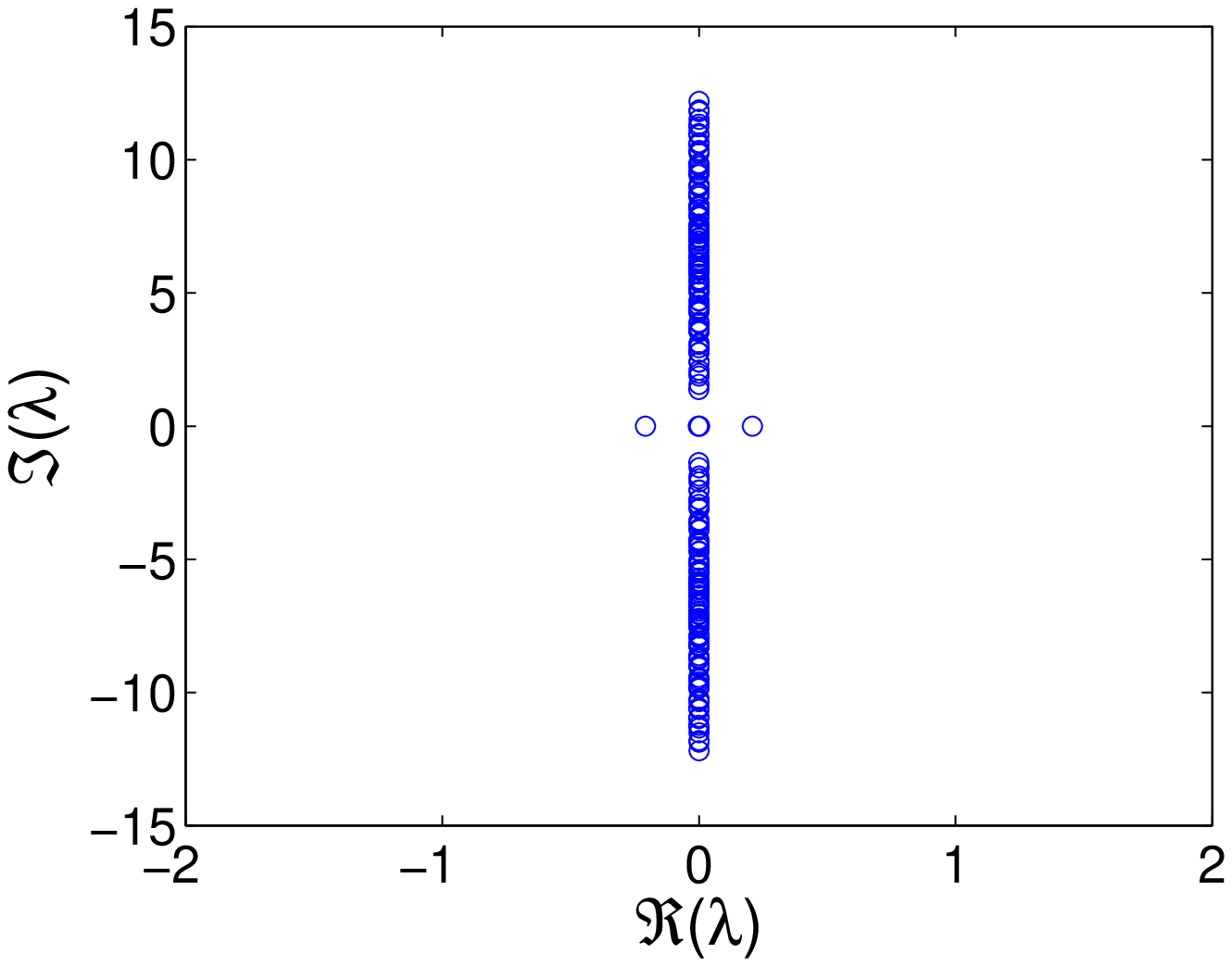}
\hskip-0.2cm
\includegraphics[width=4.4cm,angle=0,clip]{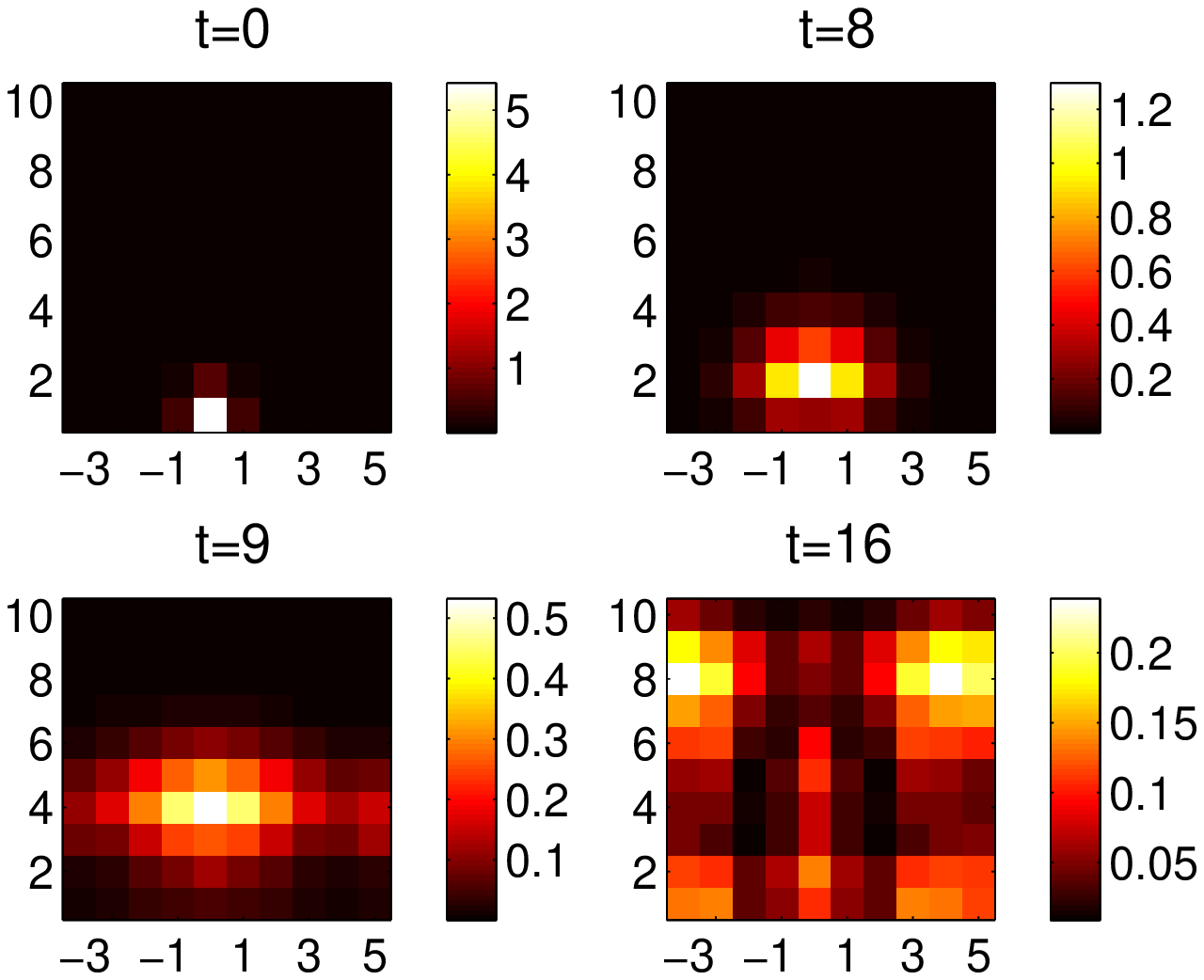}
}
\caption{(Color online) Dynamical features of the fundamental surface
soliton. The top left and right panels show, respectively, the soliton's
norm, $P$,
and the real part of the critical stability eigenvalue, versus the lattice coupling, $C$. For
comparison, the dash-dotted lines show respective quantities for a
fundamental soliton in the uniform lattice. The instability is due
to an eigenvalue pair
bifurcating from the edge of the phonon band, that eventually
crosses the origin of the spectral plane and becomes real.
In the bottom panels, we display the linear-stability spectrum of a
single-site soliton, and snapshots of its evolution (contour plots of $%
|u_{m,n}|^{2}$), for $C=1.43$, slightly above the instability threshold.
}
\label{Fig1}
\end{figure}

\section{Numerical results}
To examine the existence and stability of the
above configurations numerically, we start with the fundamental onsite
soliton at the surface, (a). Basic results for this state are displayed in
Fig.~\ref{Fig1}. At $C=0$, there is a double zero eigenvalue due to the
phase invariance. For small $C>0$, this is the only eigenvalue of the
linearization near the origin of the spectral plane
$(\Re(\lambda),\Im(\lambda))$.
As $C$ increases,
one encounters a critical value, at which an additional (but still
marginally stable)
eigenvalue
bifurcates from the edge of the continuous spectrum, as mentioned above.
With the further increase of $C$,
this bifurcating eigenvalue arrives at the origin of the spectral
plane, and subsequently gives
rise to an \emph{unstable} eigenvalue pair,
with $\Re(\lambda )\neq 0$, see Fig.~\ref{Fig1}.
This happens for $C>1.41$; we note in passing that the
results reported herein have been obtained for lattices of
size $10 \times 10$, but it has been verified that a similar
phenomenology persists for larger lattices of up to $25 \times 25$.
For comparison, we also display, by a dashed-dotted line, the
critical unstable eigenvalue for a fundamental soliton on the uniform
lattice (as a matter of fact, for a soliton sitting far from the
edge), which demonstrates that the interaction with the edge leads to
conspicuous \textit{expansion of the stability
interval} of the fundamental soliton. This may also be justified
intuitively, as the instability of the fundamental soliton emerges
closer to the continuum limit which is well-known to be unstable
to (very slow) collapse type phenomena. A surface variant of the
relevant structure enjoys the company of fewer neighboring sites
and hence is ``more slowly'' (in parameter space) approaching
its continuum limit, in comparison with its infinite lattice sibling.

Development of the instability of the fundamental surface soliton (in the
case when it is unstable) was examined in direct simulations of
Eq.~(\ref{bulk}). As seen in Fig.~\ref{Fig1},
in this case the excitation propagates
away from the edge, expanding into an apparently disordered state.
This is justifiable as for this parametric regime there is no stable
localized state neither in the vicinity of the surface or in the
(more unstable) bulk of the uniform lattice.

\begin{figure}[tb]
\centerline{
~~
\includegraphics[width=2.0cm,height=2.0cm,angle=0,clip]{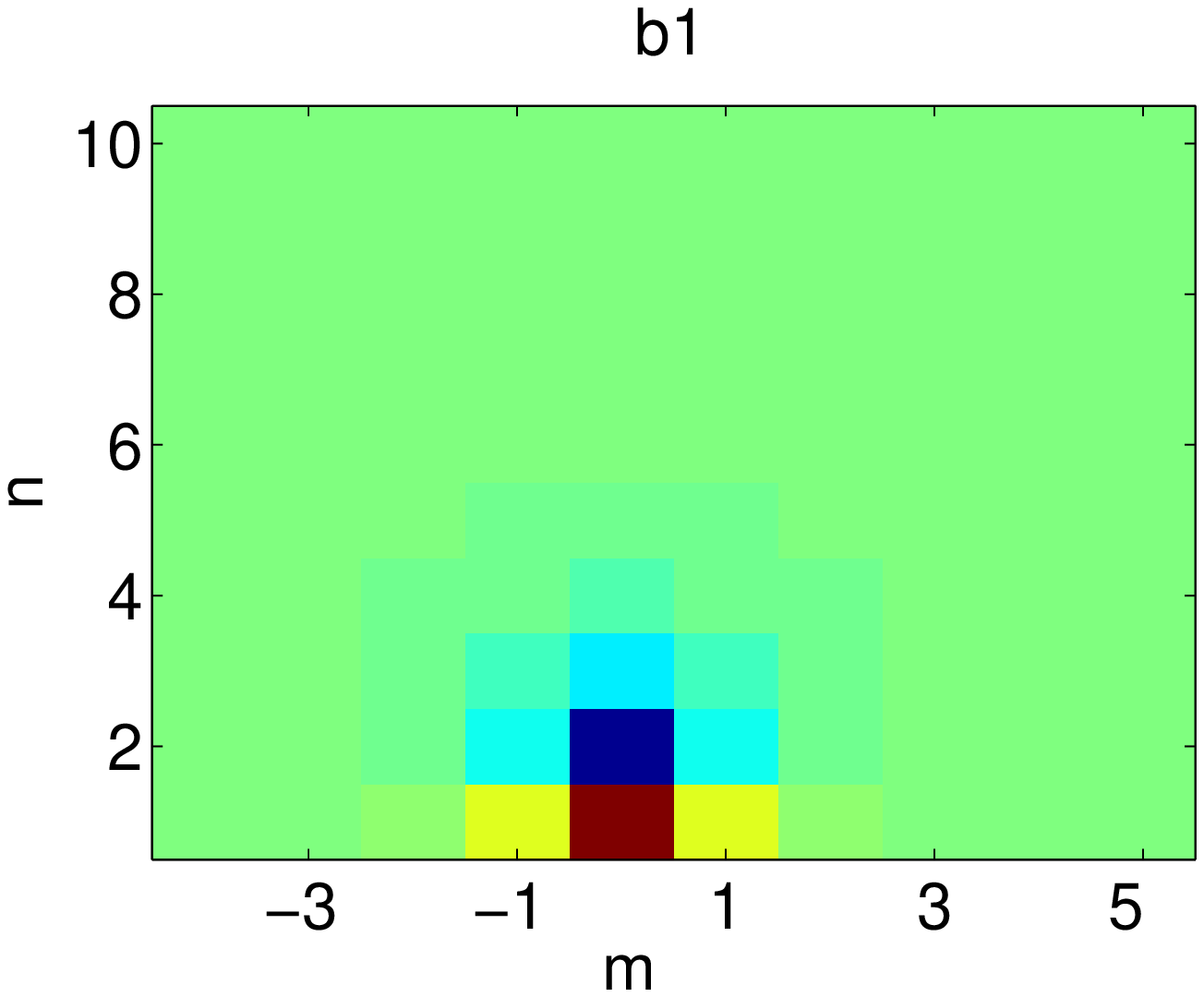}
\hskip-0.2cm
\includegraphics[width=2.0cm,height=2.0cm,angle=0,clip]{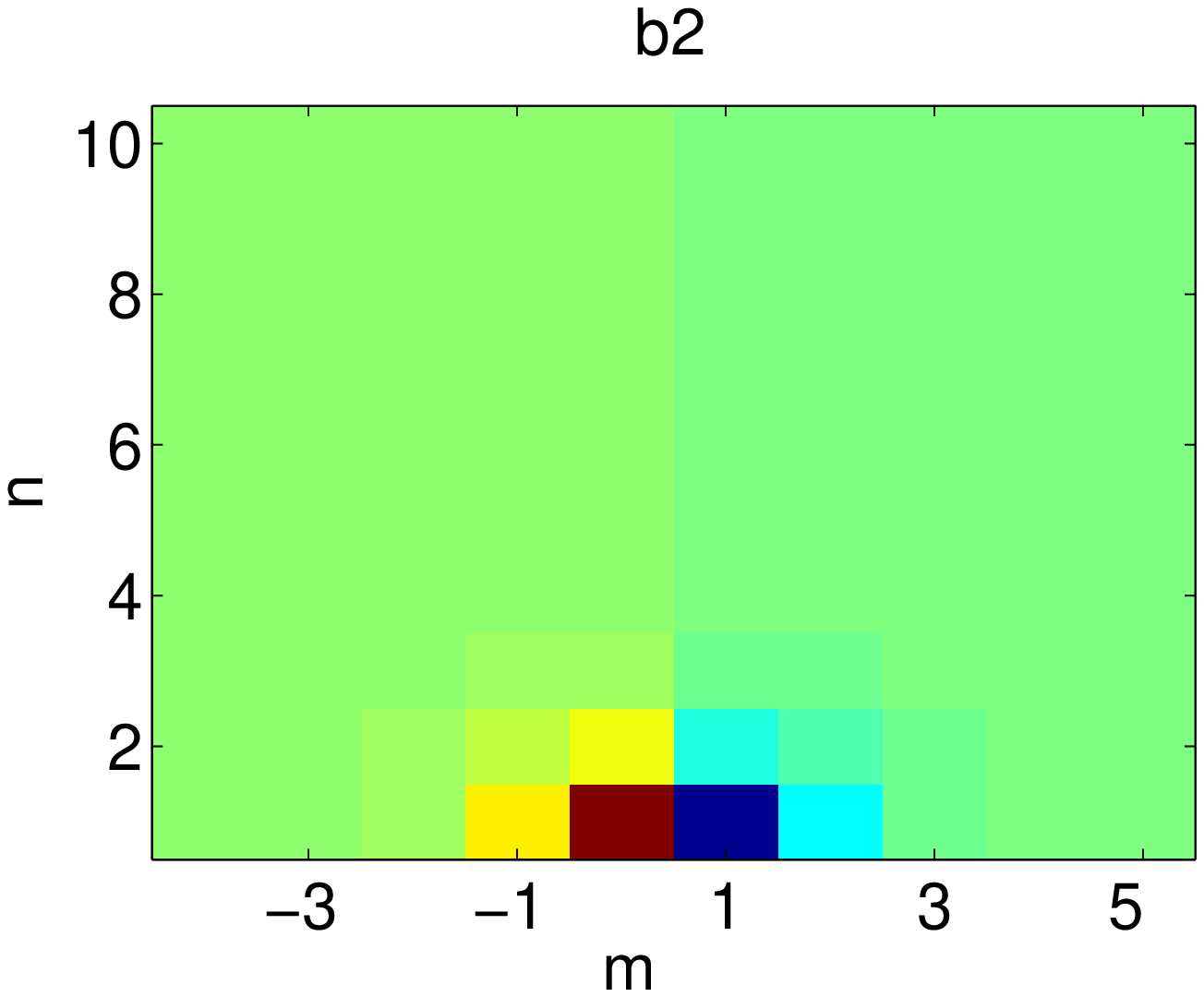}
\hskip-0.0cm
\includegraphics[width=4.4cm,angle=0,clip]{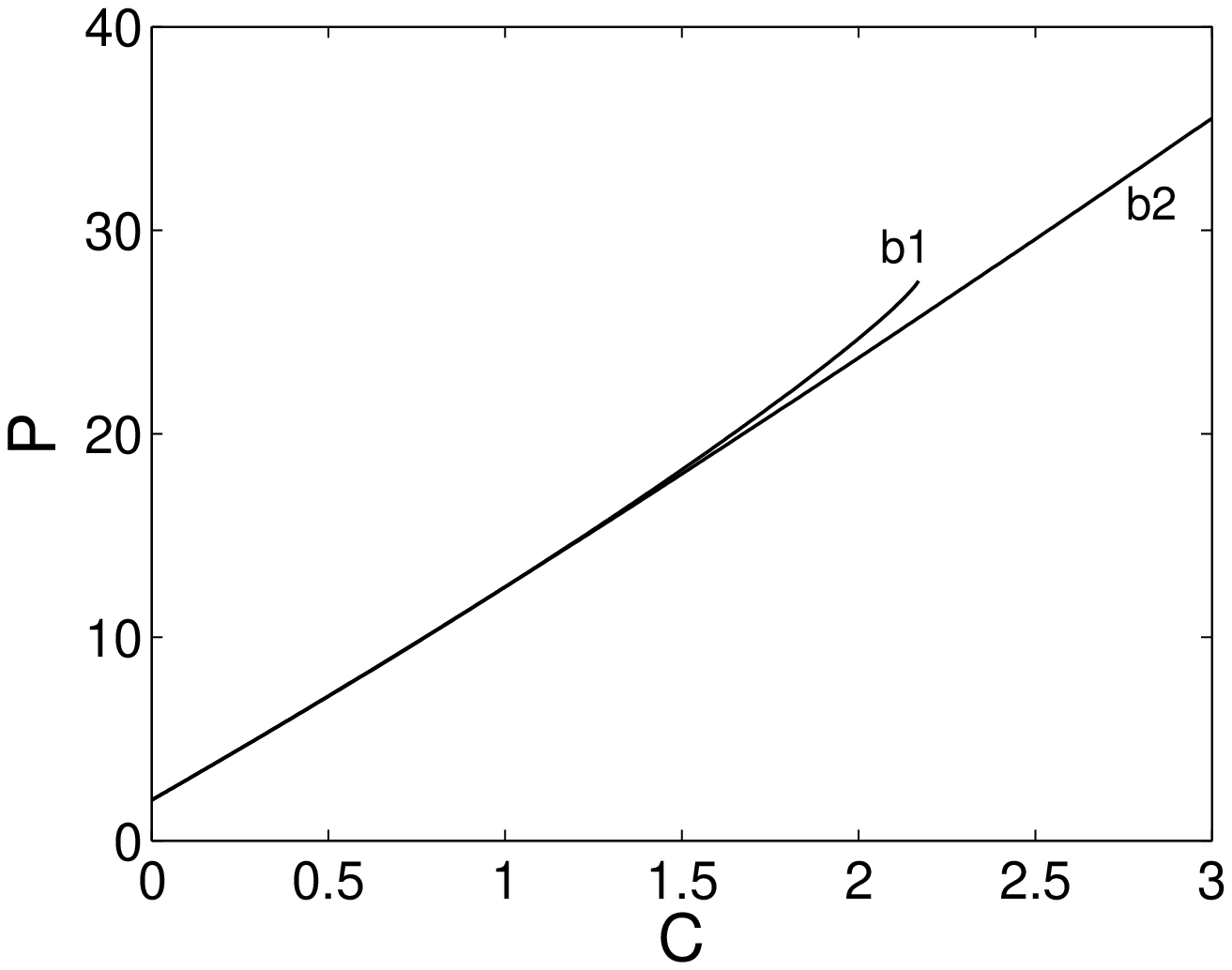}
}
\centerline{
\includegraphics[width=4.4cm,angle=0,clip]{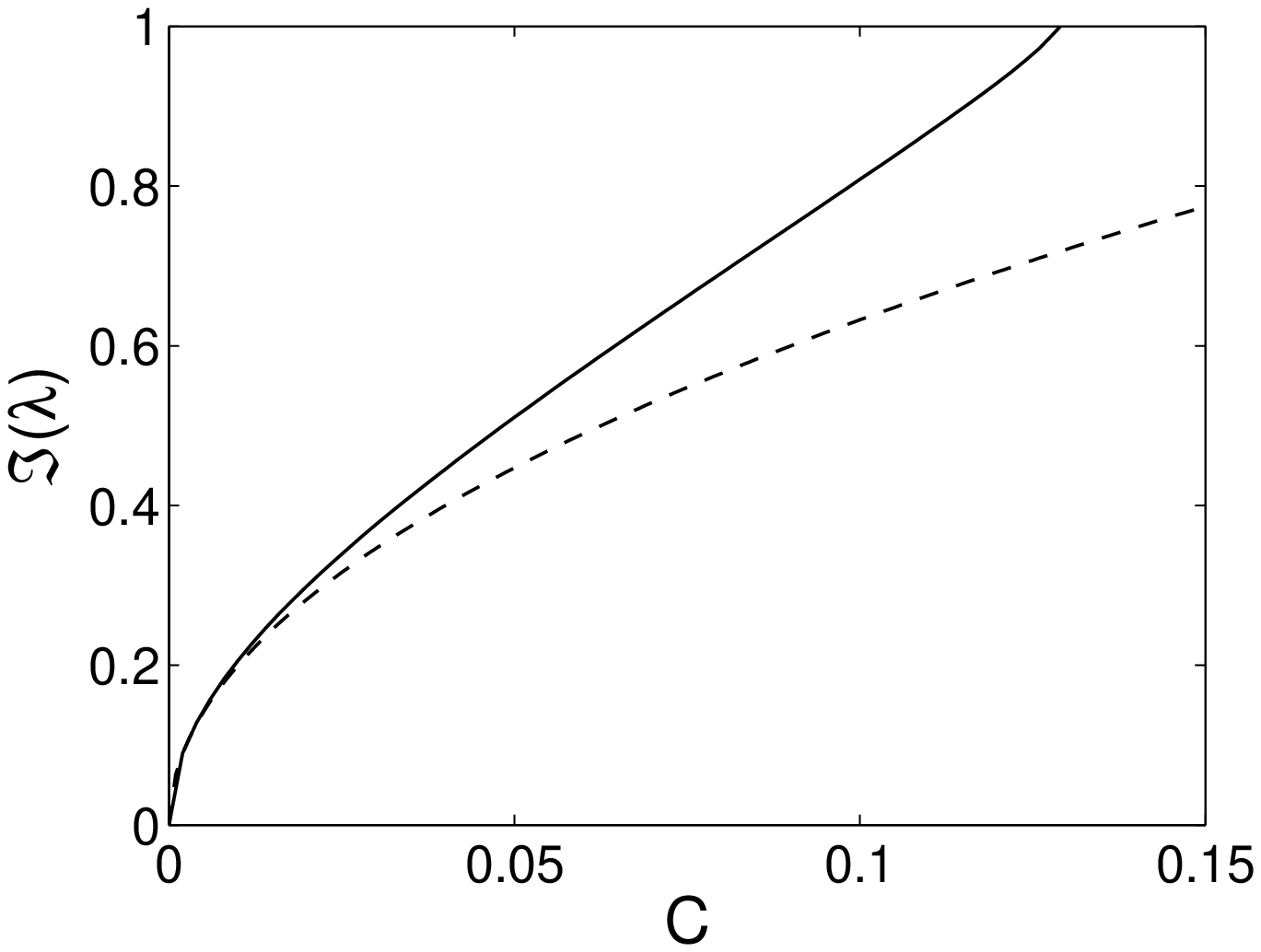}
\hskip-0.2cm
\includegraphics[width=4.4cm,angle=0,clip]{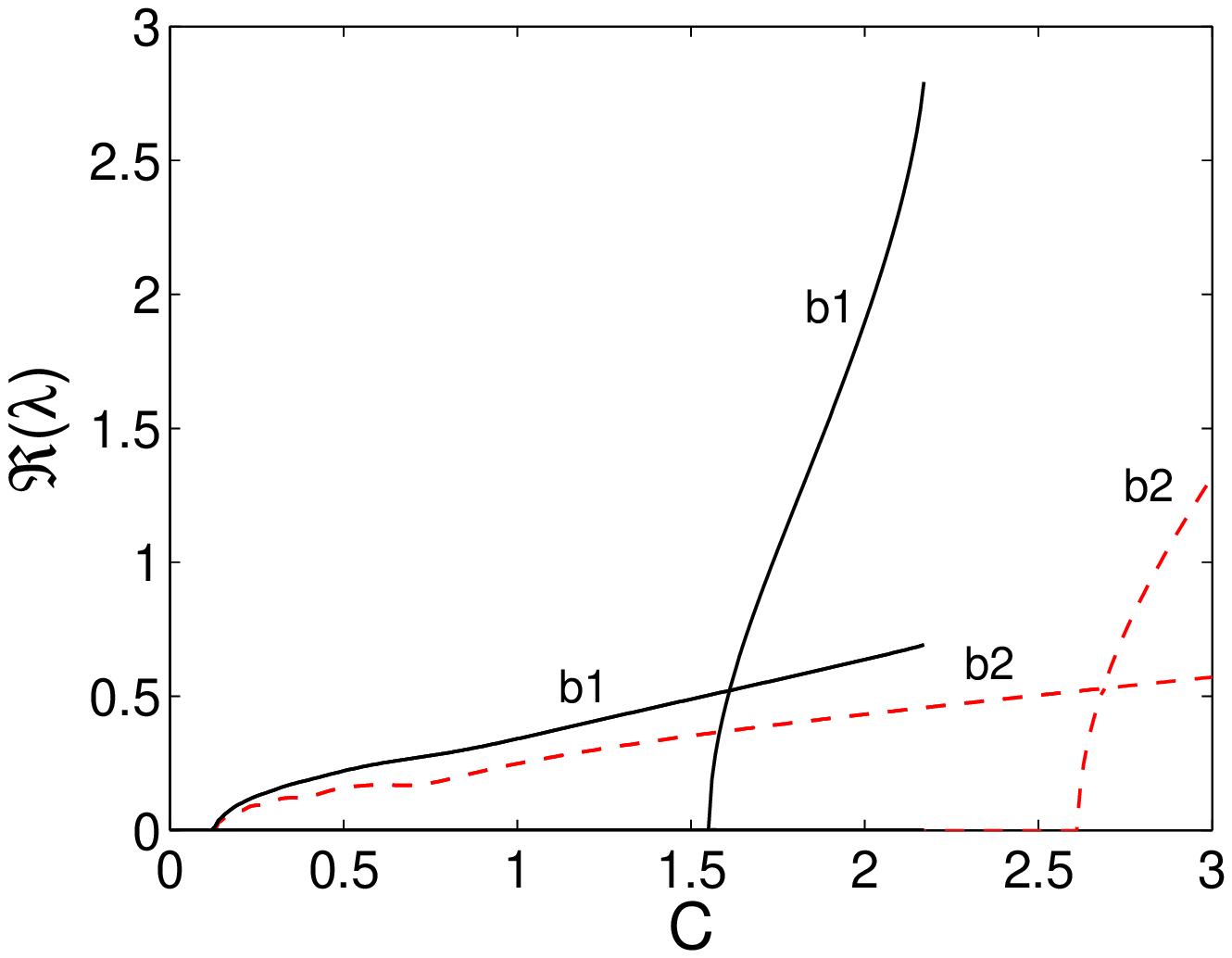}
}
\centerline{
\includegraphics[width=4.4cm,angle=0,clip]{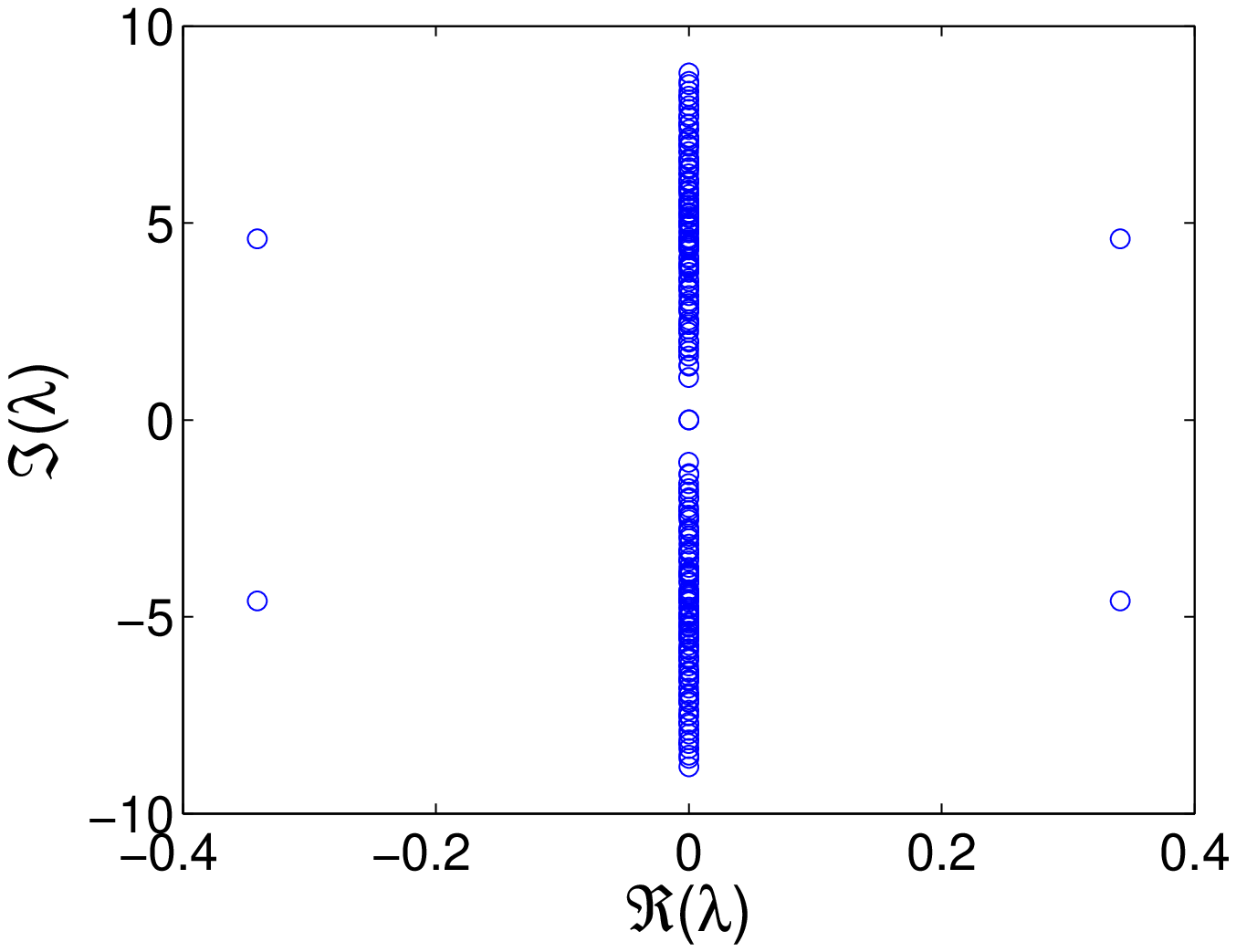}
\hskip-0.2cm
\includegraphics[width=4.4cm,angle=0,clip]{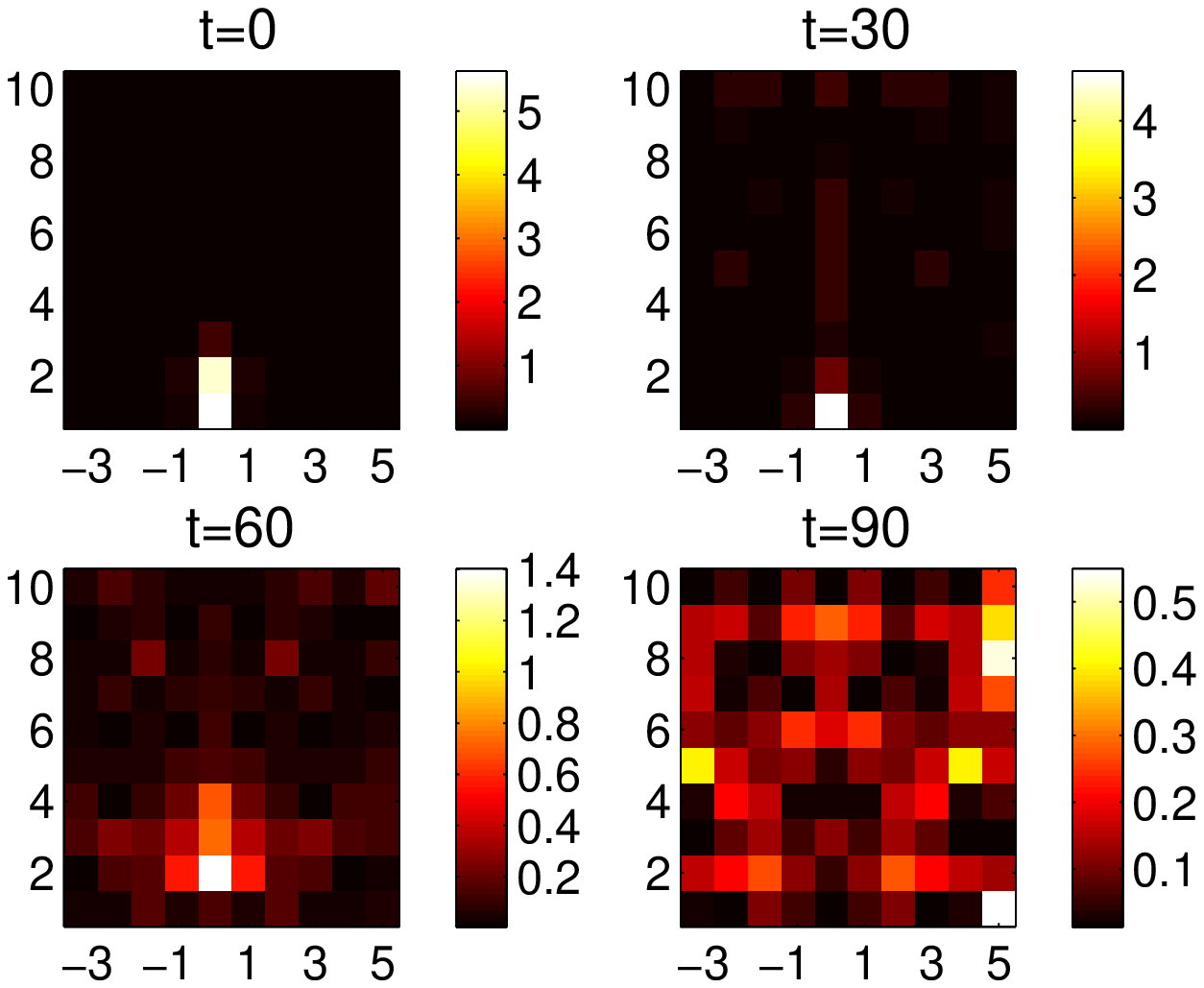}
}
\caption{ (Color online)
Top panels:
Vertical (b1) and horizontal (b2) dipoles,
which, at $C=0$, are seeded through Eq.~(\ref{dipole}).
The top left and
middle panels show examples of these dipoles at $C=1$,
and the right panel shows their norms.
Middle and bottom panels: The same as Fig.~\ref{Fig1}, but for the
configurations b1 and b2.
The vertical dipole, b1, disappears via a saddle-node bifurcation
at $C\approx
2.17$. The left middle panel displays the eigenvalue bifurcating
from zero
at $C=0$, the dashed line being the analytical approximation described in
the text, i.e., $\Im(\protect\lambda )=2\protect\sqrt{C}$. The right
middle panel shows the onset of the b1 (solid lines) and
b2 (dashed lines) instabilities for the
dipole modes, as found from numerical computations. An example of the
spectrum and nonlinear evolution of an unstable vertical dipole are
presented in the bottom panels, for $C=1$.}
\label{Fig2}
\end{figure}

Next, in Fig.~\ref{Fig2} we present results for the vertical and
horizontal dipoles,
(b1) and (b2) (see top left and middle panels of
Fig.~\ref{Fig2}, seeded as per Eq.~(\ref{dipole}).
At $C=0$, the dipole has two pairs of zero eigenvalues, one of which becomes finite (remaining
stable) for $C>0$, as shown above in the analytical form. Our numerical
findings reveal that, in compliance with the analytical results, the dipoles
of both types give rise to virtually identical finite eigenvalues (hence
only one eigenvalue line is seen in the left middle panel of Fig.~\ref{Fig2}).
As seen in the right middle panel in Fig.~\ref{Fig2}, both dipoles lose
their stability simultaneously, at $C\approx 0.15$. Continuing the
computations past this point, we conclude that the vertical and horizontal
dipoles become different when $C$ attains values $\sim 1$. Eventually, the
(already unstable) vertical configuration, b1, disappears in a saddle-node
bifurcation at $C\approx 2.17$, while its horizontal counterpart, b2,
persists through this point. Furthermore,
there is a critical value of $C$ at
which an eigenvalue
bifurcates from the edge of the continuous spectrum. Eventually,
this bifurcating eigenvalue crosses the origin of the spectral plane,
giving rise to an unstable eigenvalue pair, with $\Re(\lambda )\neq 0$.
The value of $C$ at which this secondary instability
sets in is essentially smaller for b1, i.e., $C\approx 1.55$, than
$C\approx 2.61$
for b2. We thus conclude that the horizontal dipole, b2, is, generally,
\emph{more robust} than its vertical counterpart, b1. This conclusion seems
natural, as the proximity to the edge stabilizes the fundamental soliton (as
shown above), and in the horizontal configuration the two sites that
constitute the dipole are located closer to the border.

Nonlinear evolution of unstable dipoles was examined too, in direct
simulations. As seen in the example (for configuration b1)
shown in the bottom panels of
Fig.~\ref{Fig2}, the instability
again results in a disordered state.

\begin{figure}[tb]
\centerline{
\includegraphics[width=4.4cm,angle=0,clip]{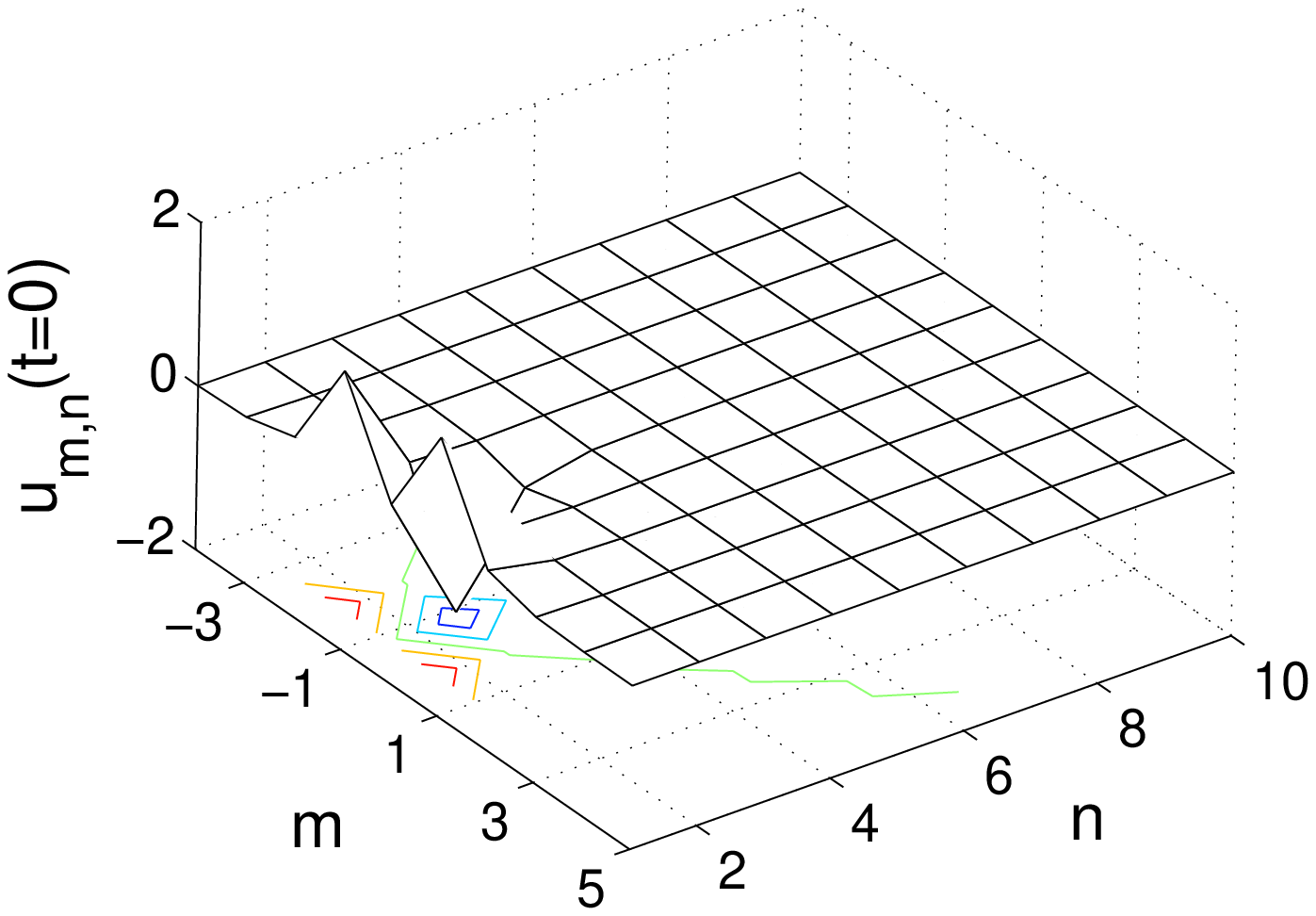}
\hskip-0.2cm
\includegraphics[width=4.4cm,angle=0,clip]{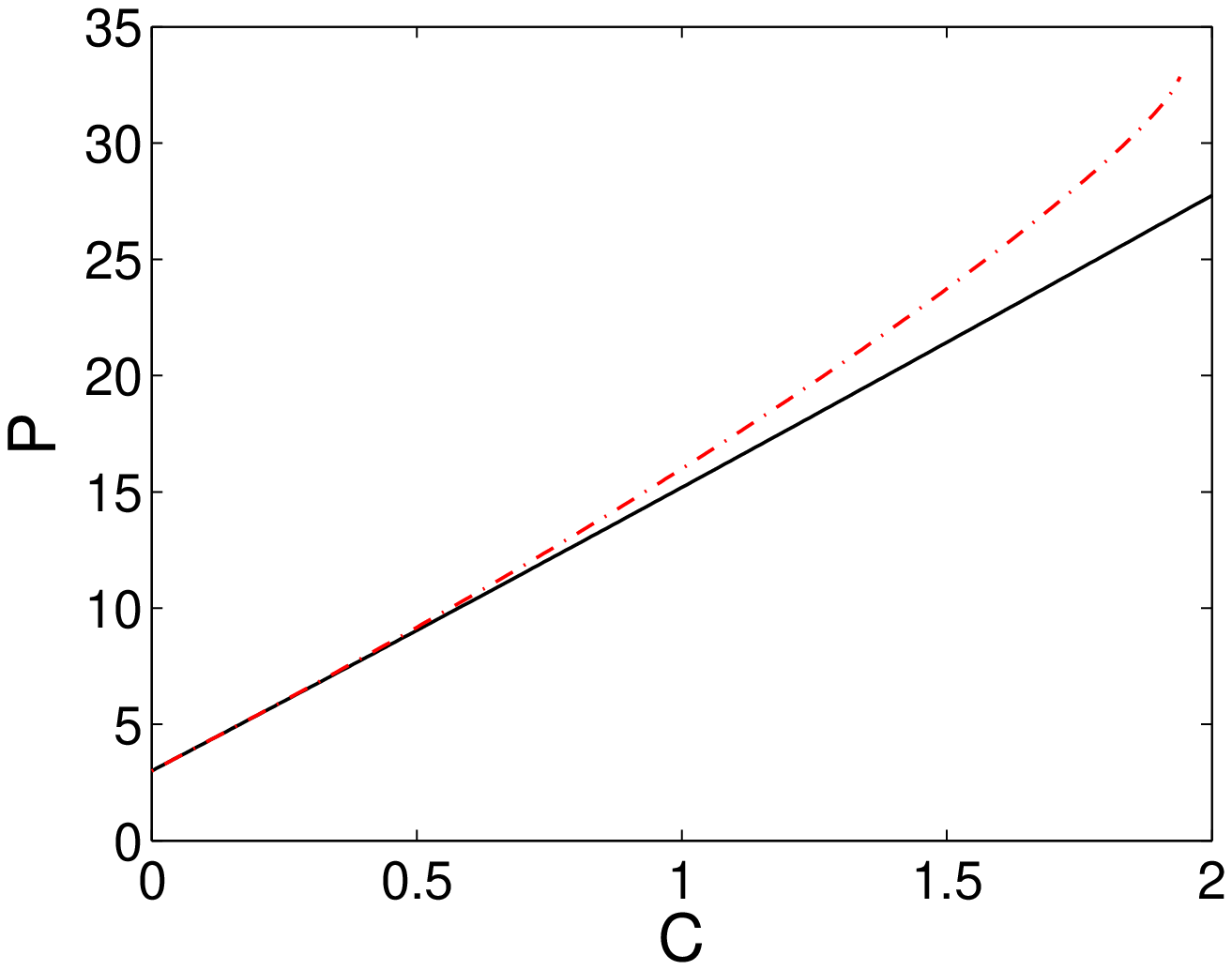}
}
\centerline{
\includegraphics[width=4.4cm,angle=0,clip]{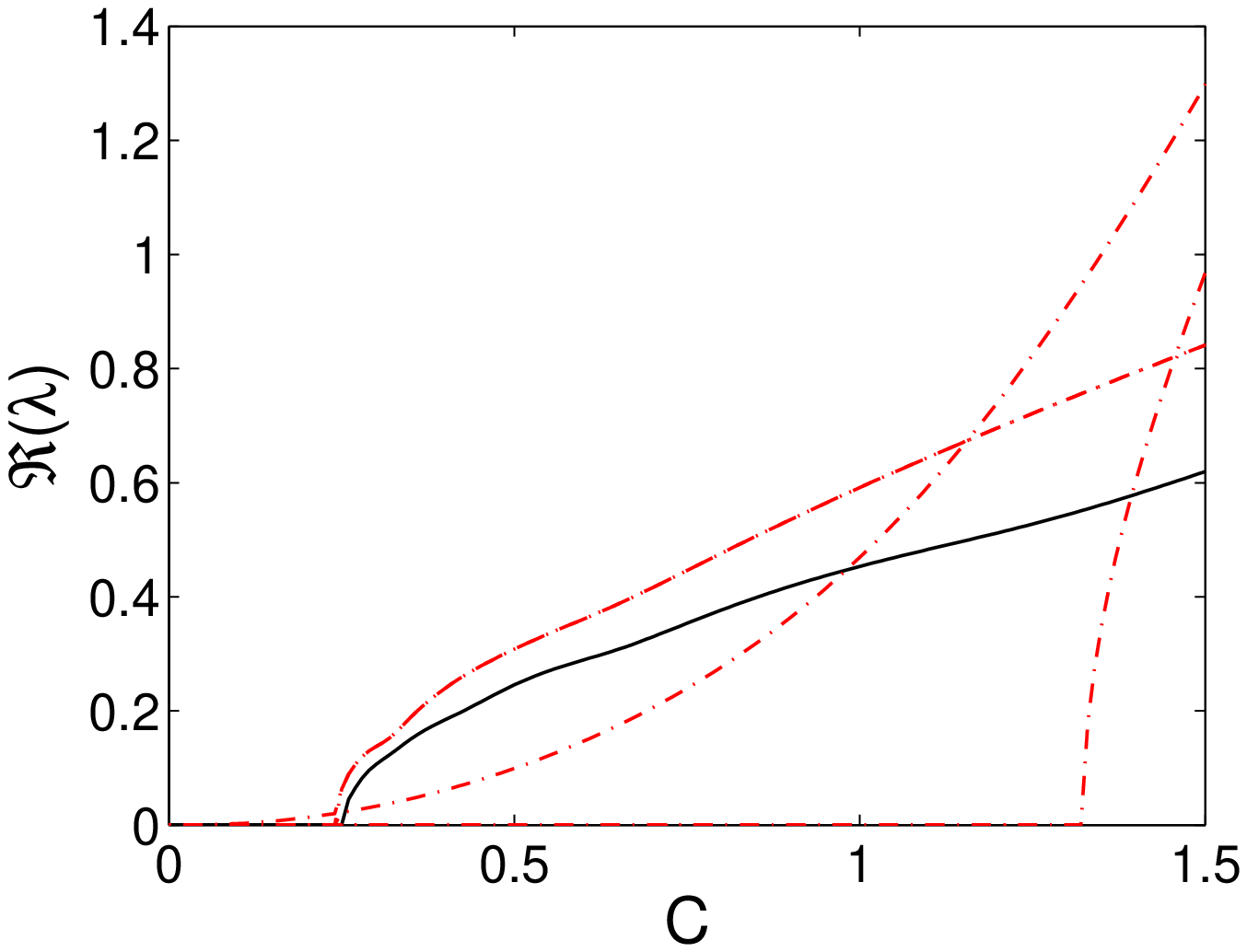}
\hskip-0.2cm
\includegraphics[width=4.4cm,angle=0,clip]{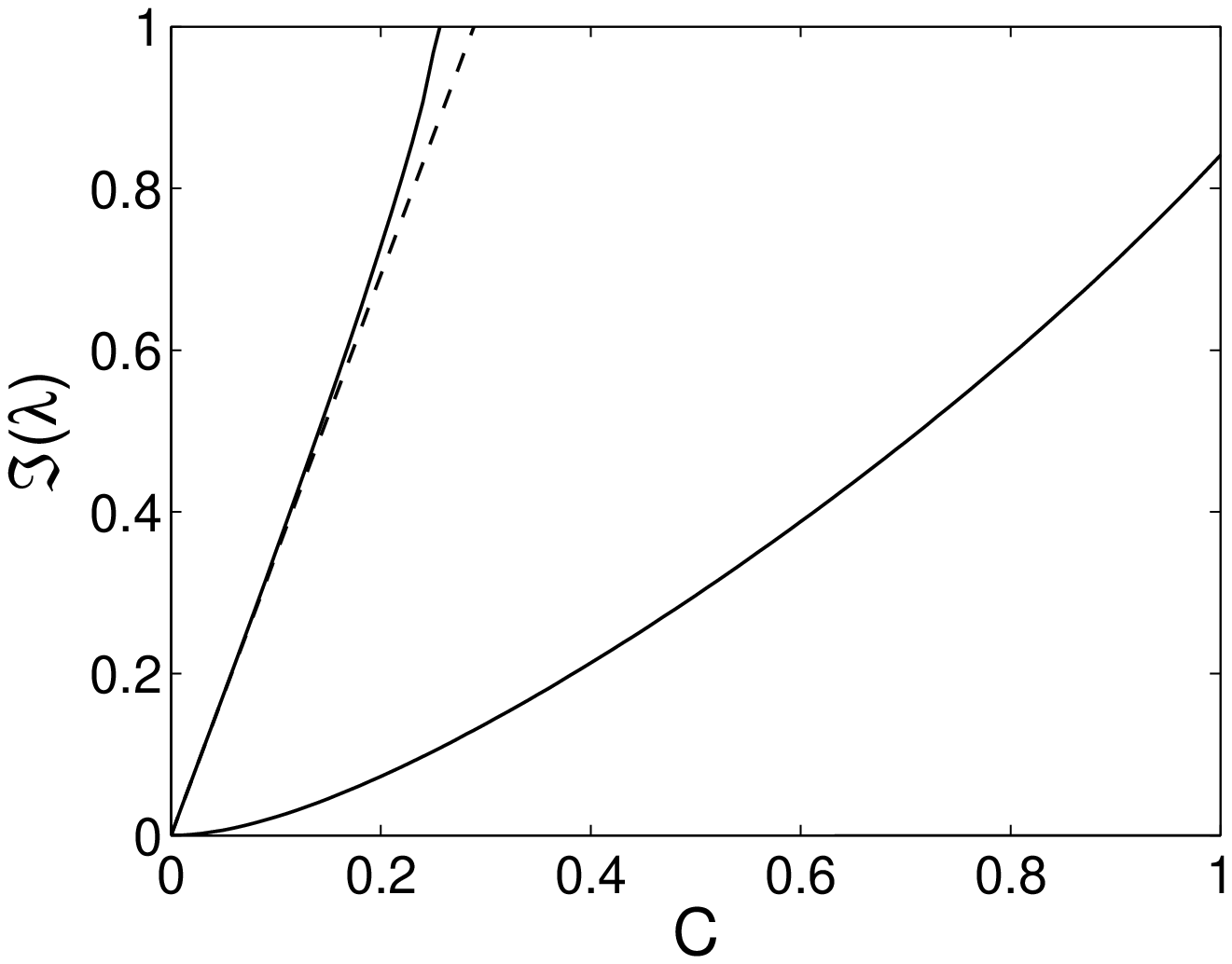}
}
\centerline{
\includegraphics[width=4.4cm,angle=0,clip]{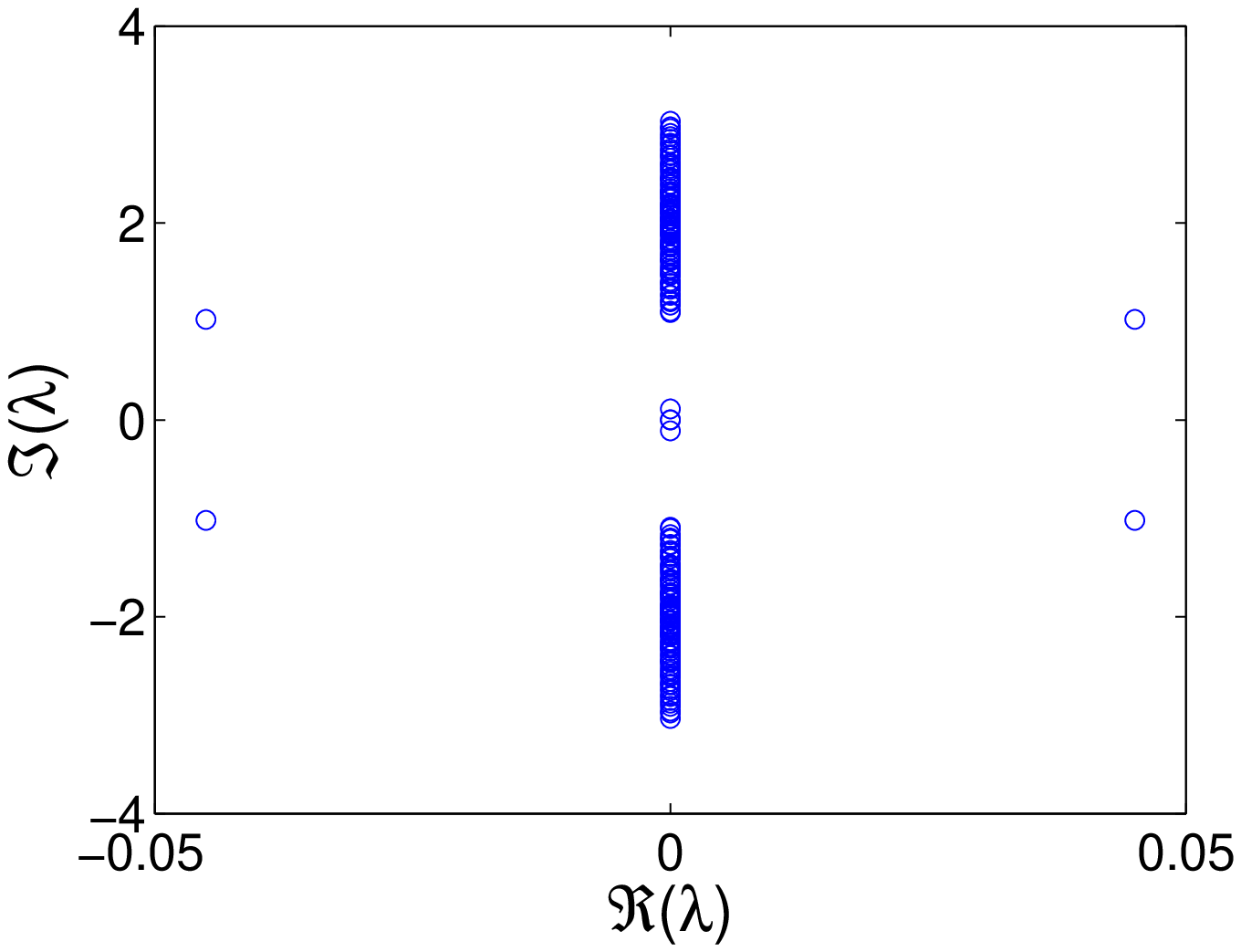}
\hskip-0.2cm
\includegraphics[width=4.4cm,angle=0,clip]{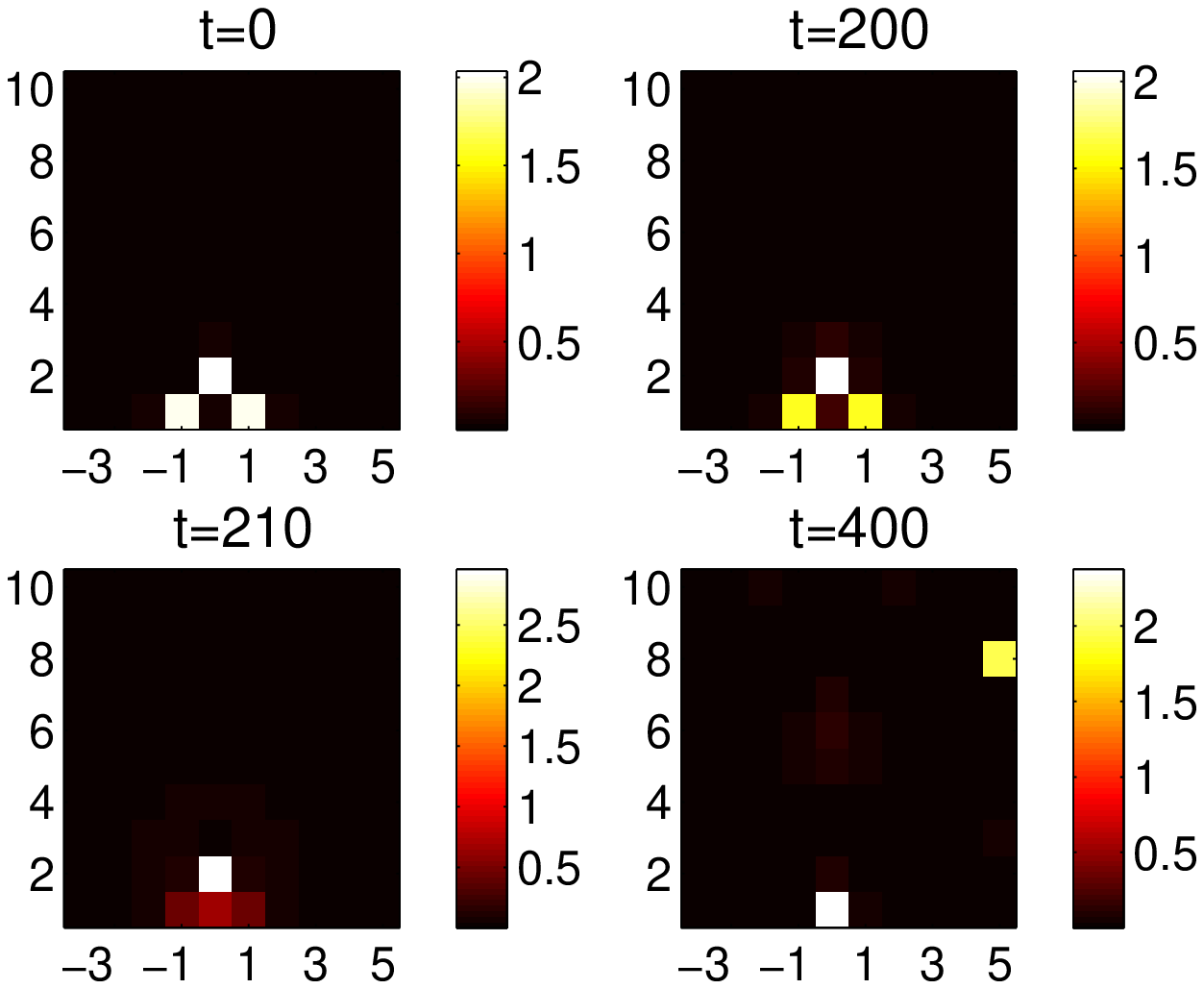}
}
\caption{(Color online) The same as Fig.~\ref{Fig2} for the
``horseshoe'' configuration, Eq.~(\protect\ref{stemcell2}).
The top panels show an example of the structure
and its norm as a function of $C$ (black solid curve).
The solid curves in the middle panels display the real and imaginary
parts of the key stability eigenvalues (the dashed line in the
right middle panel is the analytical approximation for the
imaginary part presented in the text). For comparison, the red
dash-dotted lines show the same characteristics for a family of
horseshoe solitons created in the uniform lattice. It is seen
that the latter family is {\it completely unstable}, while the
same waveform trapped at the edge of the lattice has a
well-defined stability region. This is a remarkable consequence
of the presence of the surface.
The bottom panels present the linear
instability spectrum of the horseshoe state with $C=0.26$, and its
evolution in time.} \label{Fig3}
\end{figure}

Proceeding to the completely novel configuration (c) of the
horseshoe,
we note that, because it was
seeded at three sites when $C=0$, there are three pairs of zero eigenvalues in
the AC limit. Above, it was shown analytically that one pair of these
eigenvalues becomes finite at order $\mathcal{O}(C)$, and another at
$\mathcal{O}(C^{2})$. Numerical results demonstrate that the first pair
remains stable until it collides with the edge of the continuous spectrum,
which happens at $C\approx 0.25$. As mentioned above, the second eigenvalue
pair, bifurcating from zero at order $\mathcal{O}\left( C^{2}\right) $, is
critical for the stability of configuration (c). The numerical results show
that this pair bifurcates into a \emph{stable} (i.e., imaginary)
one, and, as shown in
Fig.~\ref{Fig3}, the horseshoe remains stable up to $C\approx 0.25$.

To understand the effect of the surface on the stability of the horseshoes,
it is relevant to compare them to their counterparts in the infinite
lattice, i.e., similar structures
that may be created, starting from the AC limit
taken as per Eq.~(\ref{stemcell2}), far from the domain edge.
By itself, the latter is a
novel family of localized solutions to the DNLS equation in 2D.
However, the important feature
is that,
the structure in the bulk of the uniform lattice
(unlike the quadrupole that may be stable), is unstable for
the \emph{entire
family} of horseshoes [in fact,
the $\mathcal{O}(C^{2})$ eigenvalue pair, bifurcating from zero at $C=0$,
immediately becomes real in this case, see the corresponding,
quadratically growing dashed-dotted line
in the left middle panel in Fig.~\ref{Fig3}]. Thus, the horseshoe attached
to the surface is another example of the localized mode stabilized by the
lattice's edge (however, unlike the above example of the stabilized
fundamental
soliton, the dependence of the horseshoe's stability on the border is
crucial, as it may never be stable in the infinite lattice).

The bottom panels of Fig.~\ref{Fig3} exemplify the evolution of the horseshoe
when it is unstable. Unlike the situations with
configurations (a) and (b), the unstable horseshoe does not decay into a disordered state, but
rather splits into a pair of two fundamental solitons (with $S=0$), one
trapped at the surface and one found deeper inside the lattice.

\begin{figure}[tb]
\centerline{
\includegraphics[width=4.0cm,angle=0,clip]{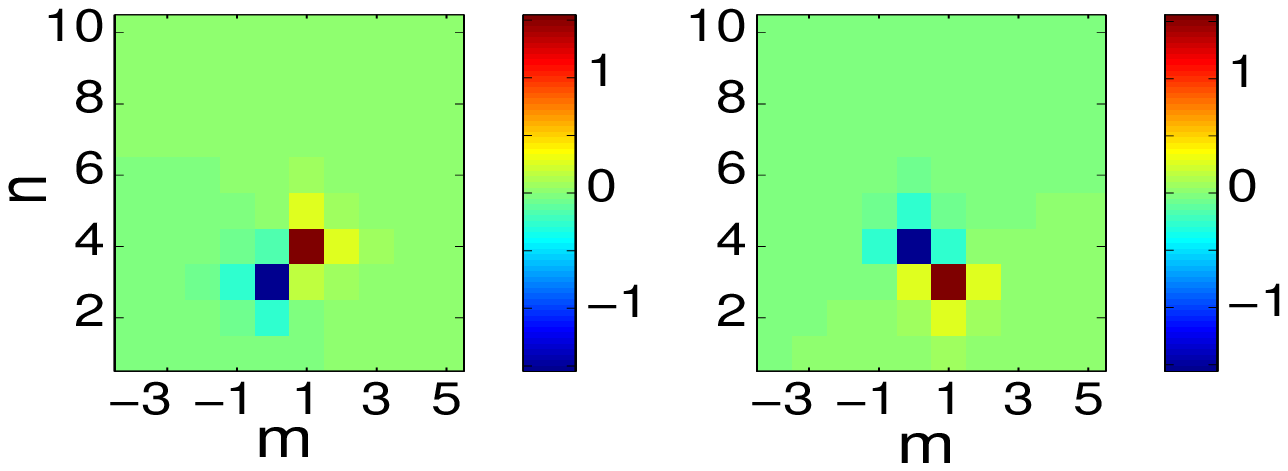}
\hskip-0.2cm
\includegraphics[width=4.4cm,angle=0,clip]{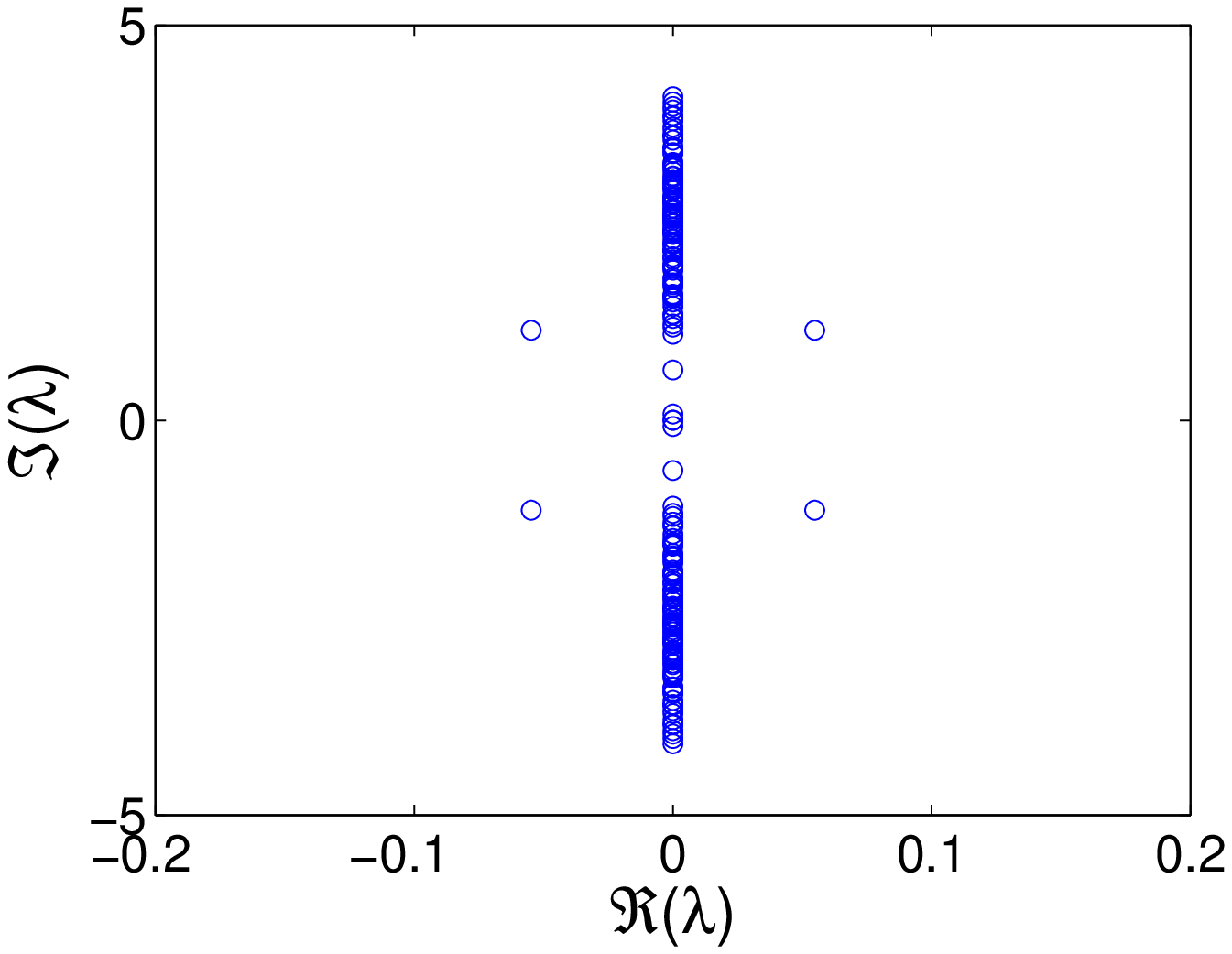}
}
\centerline{
\includegraphics[width=4.4cm,angle=0,clip]{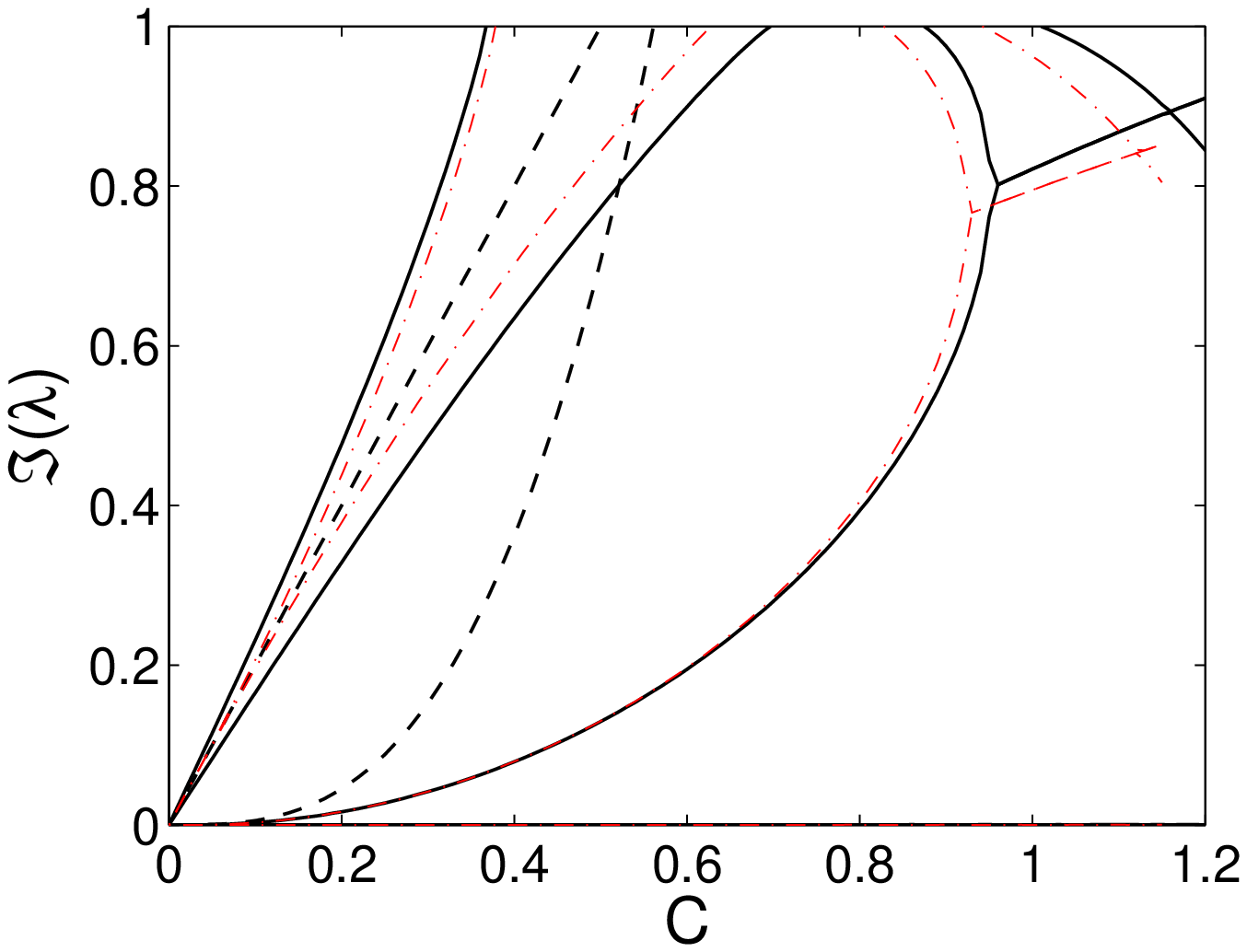}
\hskip-0.2cm
\includegraphics[width=4.4cm,angle=0,clip]{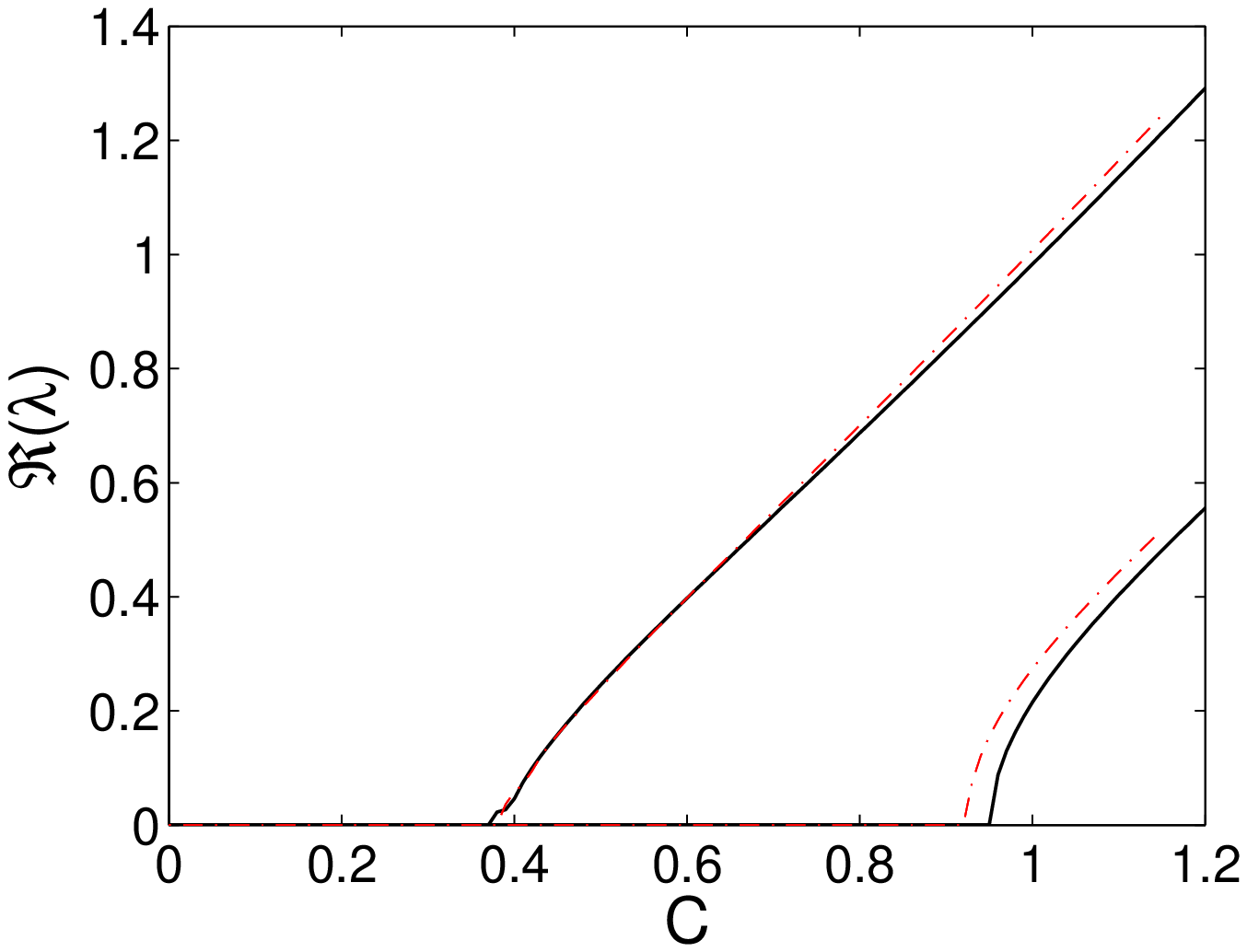}
}
\caption{(Color online) The \textit{supersymmetric} vortex cell seeded as
per Eq.~(\protect\ref{yes}). The top left and right panels show,
respectively, the real and imaginary parts of the solution and its
(in)stability spectrum for $C=0.4$. The bottom panel displays imaginary and
real parts of the stability eigenvalues versus $C$. The solid and dashed
lines show numerical and analytical results for small $C$. For comparison,
red dashed-dotted lines depict the same numerically found characteristics
for a supersymmetric vortex on the infinite lattice.}
\label{Fig4}
\end{figure}

\section{Surface effect on the existence of vortices}
Besides
the stabilization effect reported above, the lattice edge may exert a
different
effect, impeding the existence of some
solutions. As an example, we consider the, so-called \textit{supersymmetric }
\cite{peli05},
lattice vortex
attached to the edge, i.e., one with the vorticity ($S=1$) equal to the size
of the square which seeds the vortex at $C=0$, through the following set of
four excited sites, cf.~Eq.~(\ref{stemcell2}):
\begin{equation}
%
\left\{v^{(0)}_{0,1},v^{(0)}_{1,1},v^{(0)}_{1,2},v^{(0)}_{0,2}\right\}
=\left\{e^{i\theta_{0,1}},e^{i\theta_{1,1}},e^{i\theta_{1,2}},e^{i\theta_{0,2}}\right\},
\label{no}
\end{equation}
with
$\theta_{0,1}=0$, $\theta_{1,1}=\pi/2$, $\theta_{1,2}=\pi$, and $\theta_{0,2}=3\pi/2$
(unlike the above configurations, this one is not
a purely real one).
While supersymmetric vortices exist in uniform lattices (including
anisotropic ones) and have their own stability regions there \cite{peli05,we3},
numerical analysis shows that the localized mode initiated as per Eq.~(\ref%
{no}) in the model with the edge \emph{cannot} be continued to
$C>0$. In fact, we have found that, to create such a state at
finite $C$, we need to seed it, at least, two sites away from the edge, i.e., as%
\begin{equation}
\left\{v^{(0)}_{0,3},v^{(0)}_{1,3},v^{(0)}_{1,4},v^{(0)}_{0,4}\right\} =
\left\{e^{i\theta_{0,3}},e^{i\theta_{1,3}},e^{i\theta_{1,4}},e^{i\theta_{0,4}}\right\}.
\label{yes}
\end{equation}
[the set on the right-hand side is a translated version  of that of
Eq.~(\ref{no})].
Numerically found stability eigenvalues for this structure are presented in
Fig.~\ref{Fig4}, along with the analytical approximation, obtained by means
of the same method as above. In all, there are 4 pairs of
analytically predicted
eigenvalues near the spectral plane origin (given the four initial seed
sites of the configuration). More specifically, these are:
\ $\lambda =0$ (associated with the phase invariance),
$\lambda =\pm 2Ci$ (a double eigenvalue pair), and $%
\lambda =\pm \sqrt{32}C^{3}i$ (a higher order eigenvalue pair).
As it can be seen in the figure,
already at this distance of two sites from the boundary,
the behavior is sufficiently close to that of an infinite
lattice.

\section{Conclusion}
We have demonstrated that properties of
fundamental localized modes in the 2D lattice with an edge may be
drastically different from well-known features in the uniform
lattice. In particular, the edge helps to increase the stability
region for the ordinary solitons, and induces a difference between
dipoles oriented perpendicular and parallel to the lattice's
border. Additionally, regular
supersymmetric vortices cannot be created too close to the border.
Most essentially, the edge
stabilizes a new species of solitons which is entirely unstable
in the uniform lattice, the so-called ``horseshoe solitons''.
In that sense, the presence of the surface may produce a very desirable
``stability broadening'' impact on a number of different solutions;
this may become especially interesting in applications related
to waveguide arrays, among other optical or soft-matter systems.

Natural issues for further consideration are horseshoes of a larger size,
and counterparts of such localized modes in 3D
lattices near the edge. In the 3D lattice, one can also consider solitons in
the form of vortex rings or cubes \cite{we4} set parallel to the border.

PGK gratefully acknowledges support from NSF-DMS-0204585,
NSF-CAREER. RCG and PGK also acknowledge support from NSF-DMS-0505663.


\begin{thebibliography}{99}

\bibitem{book} R. K. Dodd, J. C. Eilbeck, J. D. Gibbon, and H. C. Morris.
\textit{Solitons and nonlinear wave equations} (Academic Press, London,
1982).

\bibitem{Gerard}
G. Maugin. \textit{Nonlinear Waves in Elastic Crystals}
(Oxford University Press: Oxford, 2000).

\bibitem{StenfloPlasma}
L. Stenflo,
Physica Scripta \textbf{T63}, 59 
(1996).

\bibitem{DemetriPrediction}
K. G. Makris, S. Suntsov,
D. N. Christodoulides, G. I. Stegeman, and A. Hache,
Opt. Lett. \textbf{30}, 2466 
(2005);
M. I. Molina, R. A. Vicencio, and Y. S. Kivshar,
Opt. Lett. \textbf{31,} 1693 
(2006).

\bibitem{Demetri} S. Suntsov, K. G. Makris, D. N. Christodoulides,
G. I. Stegeman, A. Hach\'{e}, R. Morandotti, H. Yang, G. Salamo,
and M. Sorel, 
Phys. Rev. Lett. \textbf{96}, 063901 (2006).

\bibitem{vectorial} I. L. Garanovich, A. A. Sukhorukov,
Y. S. Kivshar, and M. Molina, 
Opt. Express \textbf{14}, 4780 
(2006).

\bibitem{Molina} M. I. Molina, Phys. Rev. B \textbf{71}, 035404 (2005);
\textit{ibid}. \textbf{73}, 014204 (2006).

\bibitem{BarcelonaGS} Y. V. Kartashov, L. Torner and V. A. Vysloukh,
Phys. Rev. Lett. \textbf{96}, 073901 (2006).

\bibitem{experiment} C. R. Rosberg, D. N. Neshev, W. Krolikowski,
A. Mitchell, R. A. Vicencio, M. I. Molina, and Y. S. Kivshar,
e-print physics/0603202.

\bibitem{chi2} G. A. Siviloglou, K. G. Makris, R. Iwanow, R. Schiek,
D. N. Christodoulides, G. I. Stegeman, Y. Min, and W. Sohler,
Opt. Exp. \textbf{14}, 5508 
(2006).

\bibitem{BarcelonaCrossVortexSaturable} Y. V. Kartashov,
A. A. Egorov, V. A. Vysloukh, and L. Torner, 
Opt. Express \textbf{14}, 4049 
(2006).


\bibitem{we} B. A. Malomed and P. G. Kevrekidis, 
Phys. Rev. E \textbf{64}, 026601 (2001).

\bibitem{photorefr} D. N. Neshev, T. J. Alexander,
E. A. Ostrovskaya, and Y. S. Kivshar, H. Martin, I. Makasyuk,
and Z. Chen,  
Phys. Rev. Lett. \textbf{92}, 123903 (2004);
J. W. Fleischer, G. Bartal, O. Cohen, O. Manela, M. Segev, J. Hudock, and
D. N. Christodoulides,
Phys. Rev. Lett. \textbf{92}, 123904 (2004).

\bibitem{Molina2D} M. I. Molina, e-print nlin.PS/0604060. 


\bibitem{Yaroslav} Y. V. Kartashov, private communication.

\bibitem{peli05} D. E. Pelinovsky, P. G. Kevrekidis and D. J. Frantzeskakis,
Physica D \textbf{212}, 20 (2005).



\bibitem{Bishop} P. G. Kevrekidis, B. A. Malomed, and A. R. Bishop,
J. Phys. A: Math. Gen. \textbf{34}, 9615 
(2001).

\bibitem{dipole} J. Yang, I. Makasyuk, A. Bezryadina,
and Z. G. Chen,
Stud. Appl. Math. \textbf{113}, 389 
(2004);
Z. G. Chen, H. Martin, E. D. Eugenieva,
J. J. Xu, and J. K. Yang,
Opt. Express \textbf{13}, 1816
 (2005).

\bibitem{we2} P. G. Kevrekidis, B. A. Malomed, Z. Chen,
and D. J. Frantzeskakis,
Phys. Rev. E {\bf 70}, 056612 (2004); H. Sakaguchi and B. A. Malomed,
Europhys. Lett. \textbf{72}, 698 
(2005).


\bibitem{we3} P. G. Kevrekidis, D. J. Frantzeskakis,
R. Carretero-Gonz\'alez, B. A. Malomed, and A. R. Bishop,
Phys. Rev. E \textbf{72}, 046613 (2005).

\bibitem{we4} P. G. Kevrekidis, B. A. Malomed, D. J. Frantzeskakis
and R. Carretero-Gonz\'{a}lez,
Phys. Rev. Lett. \textbf{93}, 080403 (2004);
R. Carretero-Gonz\'{a}lez, P. G. Kevrekidis, B. A.
Malomed and D. J. Frantzeskakis,
\textit{ibid}. \textbf{94}, 203901 (2005).

\end{thebibliography}
\end{document}